\documentclass[amssymb,prb,twocolumn,showpacs]{revtex4}
\usepackage{epsfig}
\usepackage{dcolumn}
\usepackage{amsmath}
\begin{document}
\title{\bf\ Microscopic theory for radiation-induced
Zero-Resistance States in 2D electron systems: Franck-Condon blockade}
\author{J. I\~narrea$^{1,2}$}
\affiliation{$^1$Escuela Polit\'ecnica
Superior,Universidad Carlos III,Leganes,Madrid,Spain\\
$^2$Unidad Asociada al Instituto de Ciencia de Materiales, CSIC,
Cantoblanco,Madrid,28049,Spain.}
\date{\today}
\begin{abstract}
We present a microscopic model on radiation-induced zero resistance
states according to a novel approach: Franck-Condon physics and blockade. Zero resistance states  rise up
from   radiation-induced   magnetoresistance oscillations
 when the light intensity is strong enough.
The theory starts off with the {\it radiation-driven electron orbit model} that
proposes an interplay of the swinging nature of the radiation-driven Landau states and
the presence of charged impurity scattering. When the intensity of radiation is high enough it
turns out that the driven-Landau states (vibrational states) involved in the scattering process are spatially far from each other and
the corresponding electron wave functions do not longer overlap.
As a result, it takes place a drastic suppression of the scattering probability and then current
and magnetoresistance exponentially drop. Finally zero resistance states rise up.
This is an application to magnetotransport in two dimensional electron systems of
 the Franck-Condon blockade, based on the Franck-Condon physics which in turn stems from  molecular vibrational spectroscopy.

\end{abstract}
\maketitle
Radiation-induced magnetoresistance ($R_{xx}$) oscillations (RIRO)\cite{mani1,zudov1} turn up in  high mobility
two-dimensional electron systems (2DES) under illumination at low
temperature ($T\sim 1K$) and  low magnetic fields ($B$) perpendicular to the 2DES.
\begin{figure}
\centering\epsfxsize=3.0in \epsfysize=3.5in
\epsffile{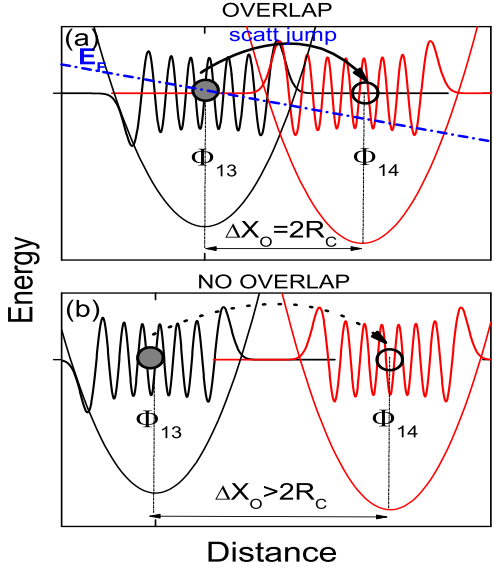}
\caption{Schematic diagrams of electron scattering between
Landau states. In Fig.(1.a) there is an important overlap
between Landau states. The case of $\Phi_{13}$ and $\Phi_{14}$ is
shown as an example. Then
the charged impurity scattering is very likely to occur.
For this to happen it is essential that the distance between
the guiding center of the Landau states is around twice the
cyclotron radius or less.
In the Fig. (1.b) we observe the opposite situation. Now the
distance is bigger than twice the cyclotron radius and the
overlap between the Landau states does not exist. Then,
the scattering process is extremely unlikely to  happen. The circles
represent the guiding center of the Landau states. }
\end{figure}
When increasing radiation power ($P$), maxima and
minima oscillations increase but the latter evolve into zero resistance states (ZRS)\cite{mani1,zudov1}.
Many experiments \cite{mani01,mani02,mani2,
mani3,mani4,smet,yuan,mani03,mani04,mani05,mani5,wiedmann1,wiedmann2,kons1,vk,mani6,mani61,mani62}
and theoretical explanations \cite{ina2,ina21,ina22,ina23,ina24,girvin,lei,ryzhii,rivera,vavilov,ina4,ina41,ina42,ina5,ina51,ina71,ina72,ryz1,ryz2,ryz3,shi}
  have been proposed to understand these effects but no consensus
among the people devoted to this field has been reached yet.
As an example of this lack of consensus, the two, in principle {\it accepted
theories}\cite{girvin,vavilov} on RIRO are no longer so relevant because they
cannot explain  basic features such as
the 1/4-cycle phase shift in the oscillation minima\cite{mani1,inaeviden} or the sublinear power law dependence
of RIRO\cite{mani6,mani62}. And they can not explain either more recent experimental evidence,
for instance, about polarization\cite{herr,ye}.
Therefore, we have to admit that to date, RIRO and ZRS are still open issues
that remain in the cutting edge of condensed
matter physics regarding radiation-mater interaction.
And this is especially true in the case of ZRS, maybe the most
intriguing and challenging  effect that shows up in this field.
Despite the fact that  plenty of theories have been developed  for RIRO,
 when it comes to ZRS only a few theoretical models have been put forward
\cite{bergeret,rivera,ina2,mikhailov,chepelianskii}.
 In general they predict negative $R_{xx}$, while it was not
experimentally confirmed.
On the other hand,  the  most accepted theory on ZRS is based on the formation of current
and electrical field domains\cite{bergeret}; the key is the
existence of an inhomogeneous current flowing through the sample
due to the presence of a domain structure.  Yet, this  is a macroscopical
model that overlooks any microscopic approach on ZRS.

In this letter  we develop a microscopic theory for ZRS that is based on
the {\it radiation-driven electron orbit model}. The model, in turn,  is based on
the exact solution of the electronic wave function in the presence
of a static magnetic field  interacting with radiation  and a perturbation treatment for
elastic scattering due to randomly distributed charged impurities.
This scattering  between Landau states, LS, (vibrational states)
 is successfully completed  when there is a net overlap between
the initial and final wave functions (see Fig. 1).
In this model the LS
semiclassically describe orbits driven by  radiation, "driven LS",
 whose center positions  oscillate according  to the radiation frequency.
This radiation-driven oscillations alter dramatically the
scattering conditions. In some cases the LS advance during the scattering jump and on average the advanced distance by
electrons is going to be bigger than in the dark giving rise to peaks in RIRO (see Fig. 2a).
In others the LS  go backward during the jump and the net distance is smaller obtaining valleys (see Fig. 2b). But in all of them
there must be a net overlap of wave functions in order to have important and valuable
contributions to $R_{xx}$.

This idea is similar to the one
in Franck-Condon physics, and extensively used in vibrational spectroscopy and
 molecular quantum mechanics\cite{levine,atkins}. ZRS turn up when the radiation intensity is high enough.
Then, it can happen that the final LS ends up behind the initial position of the scattering jump.
Although this process corresponds to a good overlap between LS, the average advanced distance is equal to zero and
does not contribute to $R_{xx}$. Then, we can consider other final LS much further with respect to the scattering initial position that could end up ahead of it
, even at very high light intensities. Nevertheless,  these LS  do not significantly overlap and the corresponding
contribution to $R_{xx}$ exponentially drops (see Fig.3). As a result, scattering rate, current and  $R_{xx}$ are
dramatically suppressed, electrons remain in their initial LS and ZRS rise up. This effect is known as
Franck-Condon blockade\cite{felix} and it is at the heart of the physical origin of ZRS

In the {\it radiation-driven electron orbits model}, the electron time-dependent Schr\"{o}dinger equation
with a time-dependent force and magnetic field
is exactly solved  to study
the magnetoresistance of a 2DES subjected to radiation at low $B$ and temperature, $T$\cite{ina2,ina21,kerner,park}.
Accordingly, the exact expression of the obtained electronic wave function reads
$\Psi_{n}(x,t)\propto\phi_{n}(x-X_{0}-x_{cl}(t),t)$,
where $\phi_{n}$ is the solution for the
Schr\"{o}dinger equation of the unforced quantum harmonic
oscillator.
Thus, the  obtained wave
function (Landau state or Landau orbit) is the same as the one of the standard quantum harmonic oscillator where the guiding
center, $X_{0}$ without radiation, is displaced by $x_{cl}(t)$.
$x_{cl}(t)$ is the classical solution of a negatively charged, forced  and damped, harmonic
oscillator:
\begin{eqnarray}
x_{cl}(t)&=&\frac{-e E_{o}}{m^{*}\sqrt{(w_{c}^{2}-w^{2})^{2}+\gamma^{4}}}\cos ( wt-\beta)\nonumber\\
&=&-A\cos ( wt-\beta)
\end{eqnarray}
$E_{0}$ the intensity of radiation, $w$ the radiation frequency and
$w_{c}$ the cyclotron frequency.
$\gamma$ is a phenomenologically introduced damping factor
for the electronic interaction with acoustic phonons.
$\beta$ is the phase difference between the radiation-driven guiding center and
the driving radiation itself and it is given by   $\tan \beta= \frac{\gamma^{2}}{w_{c}^{2}-w^{2}}$.
Thus, the guiding center lags behind radiation a phase
constant of $\beta$. When the damping parameter $\gamma$ is important, ($\gamma > w\Rightarrow \gamma^{2}>> w^{2}$), then $\tan \beta\rightarrow \infty$
 and $\beta\rightarrow \frac{\pi}{2}$.
Now, the time-dependent guiding center is, $X=X_{0}+x_{cl}=X_{0}-A \sin wt$.
This physically implies that the   orbit guiding centers oscillate harmonically at the radiation frequency $w$,
but radiation leads the guiding center displacement in $\frac{\pi}{2}$.
\begin{figure}
\centering \epsfxsize=3.0in \epsfysize=3.5in
\epsffile{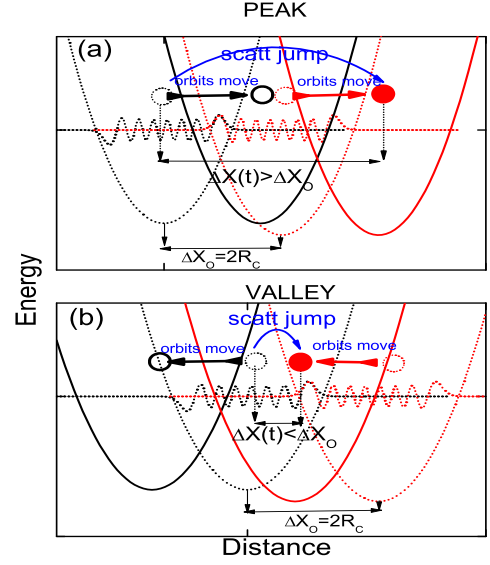}
\caption{Schematic diagram for scattering between
Landau states in the presence of radiation. Under
radiation the Landau states are harmonically driven in a swinging motion
with the radiation frequency.
In Fig. 2.a the Landau states move forward and on average
the electrons advance further than the dark case (peaks).
In Fig. 2.b the Landau states  move backward
and on average the electrons advance less than in the dark case (valleys).
For both panels dotted parabolas represent the initial driven Landau states and the
solid ones the final states after the scattering event. The circles represent
the corresponding guiding center positions of the Landau states before (dotted) and after
(solid) scattering. }
\end{figure}
\begin{figure}
\centering \epsfxsize=3.0in \epsfysize=3.5in
\epsffile{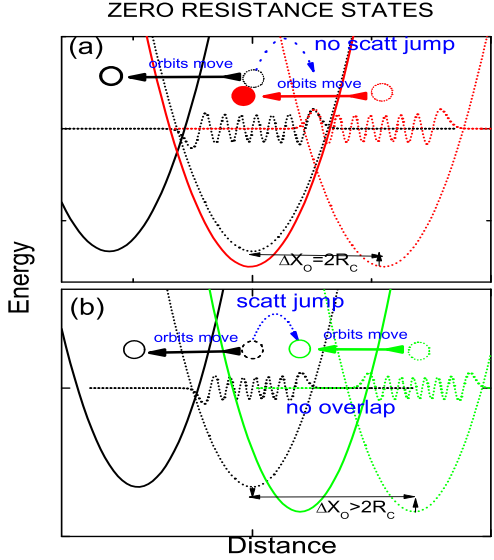}
\caption{Schematic diagram explaining  the physical origin of ZRS.
In Fig. 3.a, when the intensity of radiation is high enough and the Landau states
move backwards it may happen that the final Landau state, initially at a
distance of twice the cyclotron radius ($2R_{c}$),  ends up behind the scattering
initial position. Now and  although the overlap is important, the average advanced distance is equal to zero. In
Fig. 3.b, scattering processes to Landau states at more than $2R_{c}$. These Landau states
end up still ahead of the initial jump position and there is
a positive advanced distance. However the overlap between these
involved Landau states is negligible and the final magnetoresistance exponentially drops and
ZRS show up. }
\end{figure}

The longitudinal conductivity $\sigma_{xx}$ in the 2DES is obtained applying the Boltzmann transport theory. With this
theory and within the relaxation time approximation, $\sigma_{xx}$ is given by the following equation\cite{ridley,ando,askerov}:
\begin{equation}
\sigma_{xx}=e^{2} \int_{0}^{\infty} dE \rho_{i}(E) (\Delta X)^{2}W_{I}\left( -\frac{df(E)}{dE}  \right)
\end{equation}
being $E$ the energy and $\rho_{i}(E)$ the density of
initial Landau states. $W_{I}$ is the remote charged impurity scattering rate, given,  according to
the Fermi's Golden Rule, by
$W_{I}=\frac{2\pi}{\hbar}|<\phi_{m}|V_{s}|\phi_{n}>|^{2}\delta(E_{m}-E_{n})$,
where $E_{n}$ and $E_{m}$ are the energies of the initial and final LS respectively.
 $V_{s}$ is the scattering potential for charged impurities\cite{ando},
$\Delta X$ is the average distance advanced by the electron between
orbits in every scattering jump in the $x$ direction and is given by\cite{drivenLS},
$\Delta X= \Delta X^{0}-A\sin (2\pi\frac{w}{w_{c}})$.
$\Delta X^{0}$ is the distance between the guiding centers of the LS
involved in the scattering event. Since all LS oscillate in phase this
distance remains constant during the driving motion and is the same
with or without radiation.

After some algebra we get to an expression for $\sigma_{xx}$\cite{miura,ihn}:
\begin{eqnarray}
\sigma_{xx}&=&\frac{e^{2} m^{*}}{\pi \hbar^{2}}  \left[\Delta X^{0}-A\sin \left(2\pi\frac{w}{w_{c}}\right)\right]^{2} \nonumber\\
&\times&  W_{I} \left[ 1+\frac{2 X_{s}}{\sinh(X_{s})}
   e^{-\frac{\pi\Gamma}{\hbar w_{c}} } \cos\left(\frac{2 \pi E_{F}}{\hbar w_{c}}\right)\right]
\end{eqnarray}
where $X_{s}=\frac{2 \pi^{2}k_{B} T}{\hbar w_{c}}$, $\Gamma$ is the Landau level width
and $E_{F}$ the Fermi energy.
To find the expression of $R_{xx}$ we use
the well-known tensorial  relation
$R_{xx}=\frac{\sigma_{xx}}{\sigma_{xx}^{2}+\sigma_{xy}^{2}}
\simeq\frac{\sigma_{xx}}{\sigma_{xy}^{2}}$, where
$\sigma_{xy}\simeq\frac{n_{e}e}{B}$, $n_{e}$ being  the electron density,  and $\sigma_{xx}\ll\sigma_{xy}$.

To apply the Franck-Condon physics to the problem of ZRS we need to properly develop the
matrix element inside the scattering rate $W_{I}$. This matrix element can be expressed as\cite{ridley,ando,askerov}:
\begin{equation}
|<\phi_{m}|V_{s}|\phi_{n}>|^{2}=\sum_{q}|V_{q}|^{2}|I_{nm}|^{2}\delta_{k^{'}_{y},k_{y}+q_{y}}
\end{equation}
where $V_{q}= \frac{e^{2}}{ \epsilon (q+q_{s})}$,
 $\epsilon$ the dielectric
constant and $q_{s}$ is the Thomas-Fermi screening
constant\cite{ando}. And the integral  $I_{nm}$ is given by:
\begin{eqnarray}
I_{nm}=\int^{\infty}_{-\infty}e^{iq_{x}x} \phi_{m}(x-X^{'}) \phi_{n}(x-X)  dx
\end{eqnarray}
where $X=X_{0}-A \sin wt$ and $X^{'}=X^{'}_{0}-A \sin wt$ are the guiding centers of  $\phi_{n}$ and   $\phi_{m}$ respectively.
Expanding the exponential in the integral in powers of $q_{x}x$:\\
 $e^{iq_{x}x}=1+iq_{x}x-\frac{1}{2}q_{x}^{2}x^{2}-...$. On the one hand and using
 a screened Coulomb potential, x is of the order of the Thomas-Fermi screening length $1/q_{s}$,  $x\sim 1/q_{s}\simeq 5.10^{-9}$ $m$
 for GaAs\cite{davies}. On the other hand, $q_{x}\sim q= 2k_{F} \sin\frac{\theta}{2}$\cite{davies} where $\theta$ is the scattering
 angle  and $k_{F}$ is the Fermi wave vector. For high mobility samples
the scattering is mainly described by long range, small angle (charged impurity) scattering. Then,
we assume that for the samples used in experiments this angle is small or very small\cite{angle}.
We have taken an average scattering angle of $\theta\leq 10^{\circ}$ and for the Fermi wave vector
$2k_{F}\simeq (3-1) \times 10^{8}$ $m^{-1}$ for a 2DES with the experimental electron density.
This gives for $q_{x}\sim 10^{6}-10^{7}$  $m^{-1}$ and then $q_{x} x\sim 10^{-3}-10^{-2} << 1$.
We therefore make a good approximation retaining only the first term in the
above expansion: $e^{iq_{x}x}\rightarrow 1$.
The final outcome is that the integral $I_{nm}$ becomes an overlap integral of
the LS involved in the scattering process:\\
$I_{nm}=\int^{\infty}_{-\infty} \phi_{m}(x-X^{'}) \phi_{n}(x-X)  dx$.
This  result implies that an important overlap
between the inital and final LS will give, through the
term $|I_{nm}|^{2}$, an intense scattering  and in turn
an intense $R_{xx}$. This principle is known in Franck-Condon physics
and extensively used
in molecular vibrational spectroscopy\cite{levine,atkins}.
We translate it now into
magnetotransport in 2DES, and calculate the square of the
vibrational overlap integral, $|I_{nm}|^{2}$, {\it the Fanck-Condon factor}.
The expression for the Franck-Condon factor (FC) reads\cite{fc}:
\begin{equation}
|I_{nm}|^{2}=\frac{n!}{m!}\left[\frac{\Delta X_{0}^{2}}{2R^{2}}\right]^{m-n}e^{-\frac{\Delta X_{0}^{2}}{2R^{2}}}\left[L_{n}^{m-n}\left(\frac{\Delta X_{0}^{2}}{2R^{2}}\right)\right]^{2}
\end{equation}
where $m \geq n$,  $R^{2}=\frac{\hbar}{eB}$ is the square of the magnetic lenth
and $L_{n}^{m-n}$ the associate Lagueere polynomials.

In Fig. 4 we exhibit the calculated FC factor versus $\Delta X_{0}$ in units of cyclotron radius ($R_{c}$) for three different
$B$, : $B=0.1, 0.2$ and $0.3 T$. For each case we also present the Landau level index for the Fermi energy and
the scattering process considered in the simulation.
We observe, as expected, that the FC factor presents important values only when $\Delta X_{0}\leq 2R_{c}$ (important overlap between LS) and
exponentially drops when $\Delta X_{0}>2R_{c}$ (negligible overlap). As in vibrational transitions in
infrared molecular spectroscopy with the spectroscopic lines, here the FC factor defines the intensity of the
 scattering. Thus, when the LS involved in the scattering event are at a distance of $2R_{c}$ or less
the FC factors (see Fig.4) and in turn $W_{I}$ give important and non-negligible values. Now,
with a not very intense radiation, the final driven LS always ends up ahead of the LS initial position of scattering giving
rise to bigger or smaller $\Delta X$: peaks or valleys respectively in $R_{xx}$.
 This is described in Fig. 2.  We can get to a totally different scenario if we further increase $P$  reaching  a
situation  where the final LS ends up behind the initial scattering jump position and
then although with an important value for the FC factor, the average
advance distance is zero. Nevertheless, we can consider further away LS at more distance than $2R_{c}$ so that
they end up, even at high $P$, ahead of the initial scattering position giving a
net advanced distance. Yet,  there is no overlap now and the FC factor turns out to be negligible.
This physical scenario corresponds to the rise up of ZRS. This situation is described
in Fig. 3.

\begin{figure}
\centering \epsfxsize=3.0in \epsfysize=3.5in
\epsffile{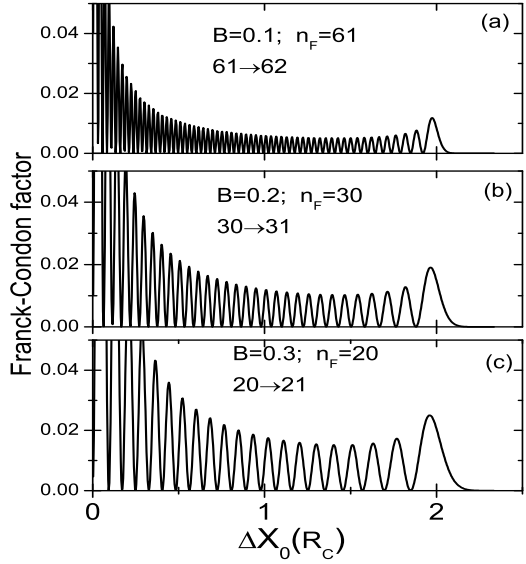}
 \caption{Franck-Condon factor vs $\Delta X_{0}$
 in units of cyclotron radius $R_{c}$for three different $B$: $B=0.1, 0.2$ and $0.3 T$. In each
panel we present also the Landau level index for the Fermi energy  and
the Landau levels indexes for the scattering process considered in the simulation. }
\end{figure}
\begin{figure}
\centering \epsfxsize=3.5in \epsfysize=3.5in
\epsffile{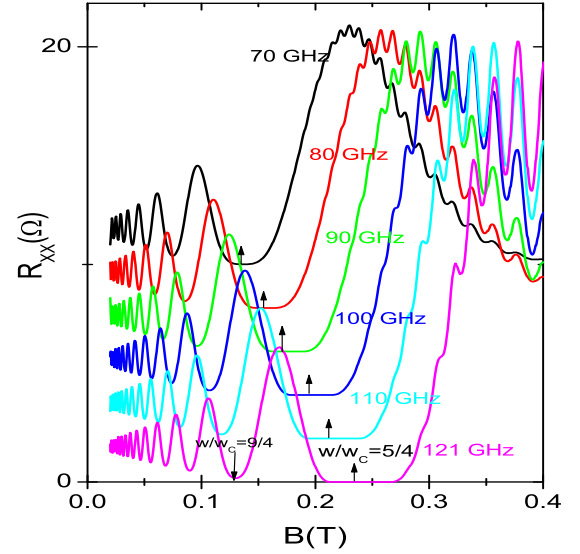}
 \caption{Calculated $R_{xx}$ vs $B$ under illumination for different radiation frequencies to study the dependence
 of radiation-induced oscillations on the frequency.}
\end{figure}
\begin{figure}
\centering \epsfxsize=3.5in \epsfysize=4.0in
\epsffile{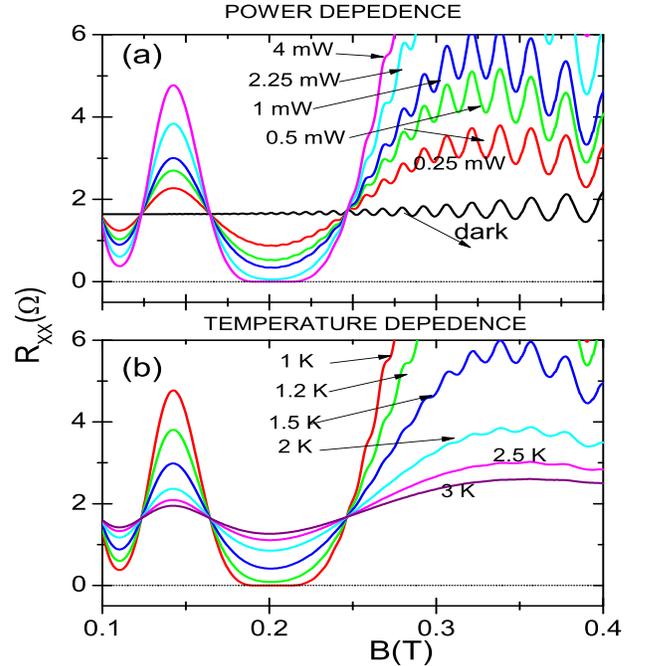}
 \caption{Calculated $R_{xx}$ vs $B$ under illumination  to study the dependence
 of radiation-induced oscillations on radiation power $P$, 6(a) and on temperature, 6(b).}
\end{figure}
 Figure 5 shows
calculated $R_{xx}$ vs $B$ for different radiation frequencies. ZRS positions move according to the changing of radiation frequency
$w$, keeping the ratio, $\frac{w}{w_{c}}=j+\frac{1}{4}$\cite{inaeviden}. Simulated ZRS are very clearly obtained for
$j=1$.
In Fig. 6, we exhibit the
dependence of $R_{xx}$ on $P$ (Fig. 6a) and on  $T$ (Fig. 6b) vs $B$ for a frequency of  $103.5$ GHz.
The rise of ZRS for both can be understood in terms of amplitude of RIRO, $A$.
In the first case, $P\propto\sqrt{E_{0}}$ and then, an increasing $P$ makes a bigger $A$
trough the radiation electric field $E_{0}$. For a certain high  $E_{0}$ we will get the
condition $\Delta X^{0}\leq A\sin (2\pi\frac{w}{w_{c}})$ and ZRS will begin to show up.
On the other hand as $T$ increases from the lowest $T=1K$, $R_{xx}$ is softened and eventually almost
disappears. The explanation can be readily obtained through the
damping parameter $\gamma$ and its influence on $A$. The damping parameter $\gamma$ is linear with $T$\cite{ina21,ando}, then
when increasing $T$ the amplitude $A$ gets smaller, wiping  out first ZRS and later RIRO.

This work is supported by the MINECO (Spain) under grant
MAT2014-58241-P  and ITN Grant 234970 (EU).
GRUPO DE MATEMATICAS APLICADAS A LA MATERIA CONDENSADA, (UC3M),
Unidad Asociada al CSIC.

\end{document}